\documentclass[12pt]{iopart}

\usepackage{cite}

\usepackage{graphicx}
\usepackage[german, english]{babel}
\usepackage{color}

\usepackage{dcolumn}
\usepackage{bm}
\usepackage{amssymb}

\begin{document}

\title{Observation of interference between two molecular Bose-Einstein condensates}

\author{C Kohstall$^{1,2}$,
    S Riedl$^{1,2}$\footnote[3]{Present address: Max-Planck-Institut f\"ur Quantenoptik, Garching, Germany.},
    E R S\'{a}nchez Guajardo$^{1,2}$,
    L~Sidorenkov$^{1,2}$,
    J Hecker Denschlag$^{1}\footnote{Present address: Institut f\"ur Quantenmaterie, Universit\"at Ulm, Germany.}$,
    and R Grimm$^{1,2}$}

\address{$^1$ Institut f\"ur Experimentalphysik und Zentrum f\"ur Quantenphysik, Universit\"at Innsbruck, 6020 Innsbruck, Austria}
\address{$^2$ Institut f\"ur Quantenoptik und Quanteninformation, \"Osterreichische Akademie der Wissenschaften, 6020 Innsbruck, Austria}

\begin{abstract}
We have observed interference between two Bose-Einstein condensates of weakly bound Feshbach molecules of fermionic $^6$Li atoms. Two condensates are prepared in a double-well trap and, after release from this trap, overlap in expansion. We detect a clear interference pattern that unambiguously demonstrates the de~Broglie wavelength of molecules. We verify that only the condensate fraction shows interference. For increasing interaction strength, the pattern vanishes because elastic collisions during overlap remove particles from the condensate wave function. For strong interaction the condensates do not penetrate each other as they collide hydrodynamically.
\end{abstract}

\maketitle

\section{Introduction}

Interference manifests the wave nature of matter. The concept of matter waves was proposed
by de~Broglie in 1923 \cite{broglie1923waq} and now represents a cornerstone of quantum physics. Already in the 1920's, experiments demonstrated the diffraction of electrons \cite{Davisson1927tso} and of atoms and molecules \cite{Estermann1930bvm}. These early achievements led to the field of atom optics and interferometry \cite{Adams1994ao,Bongs2004pwc,Cronin2009oai}.

With the realization of Bose-Einstein condensates (BECs)
\cite{Anderson1995oob,Davis1995bec,Bradley1995eob}, sources of macroscopically coherent matter waves became available.
The interference between two BECs was first observed by Andrews et
al.~\cite{Andrews1997ooi}. This landmark experiment evidenced
interference between two independent sources and revealed the relative phase between them \cite{Castin1997rpo}.
Since then, interference measurements have developed into an indispensable tool for research on BEC. Applications include detection of the phase of a condensate in expansion \cite{Simsarian2000itp}, investigation of a condensate with
vortices \cite{Inouye2001oov}, and studies of quasi-condensates \cite{hadzibabic2006bkt} or Luttinger liquids \cite{Hofferberth2007nec} in reduced  dimensions.
Another fundamental line of research in matter-wave optics is to explore the transition from the quantum to the classical world by detecting the wave nature of progressively larger particles, like clusters \cite{Schoellkopf1994nms}, C$_{60}$ \cite{Arndt1999wpd}, and other
giant molecules \cite{Gerlich2011qio}.

The creation of molecular Bose-Einstein condensates (mBECs) of
paired fermionic atoms
\cite{Jochim2003bec,Greiner2003eoa,Zwierlein2003oob} provides us with
macroscopically coherent molecular matter waves. In this article, we present the
interference of two such mBECs and demonstrate interference as a tool to investigate condensates of atom pairs. This work extends the interference of condensates towards larger, composite particles.

In a Young-type interference experiment, we release two mBECs
from a double-well trap and, after the condensates have overlapped,
we observe an interference pattern by absorption imaging.
In Sec.~\ref{sec:exp}, we describe the experimental procedures in detail.
In Sec.~\ref{sec:res}, we present our main experimental results, demonstrating the {\it molecular} de~Broglie wavelength and the dependence of the interference contrast on temperature and interaction strength.
Increasing the interaction strength reduces the visibility because of increasing elastic scattering losses depleting the coherent matter wave. Section~\ref{sec:con} gives an outlook to possible extensions and applications of interference of pair condensates.

\section{Experimental procedures}
\label{sec:exp}

\subsection{Preparation of the molecular Bose-Einstein condensate}
\label{sec:pre}

\begin{figure}
\begin{center}
 \includegraphics[width=12cm]{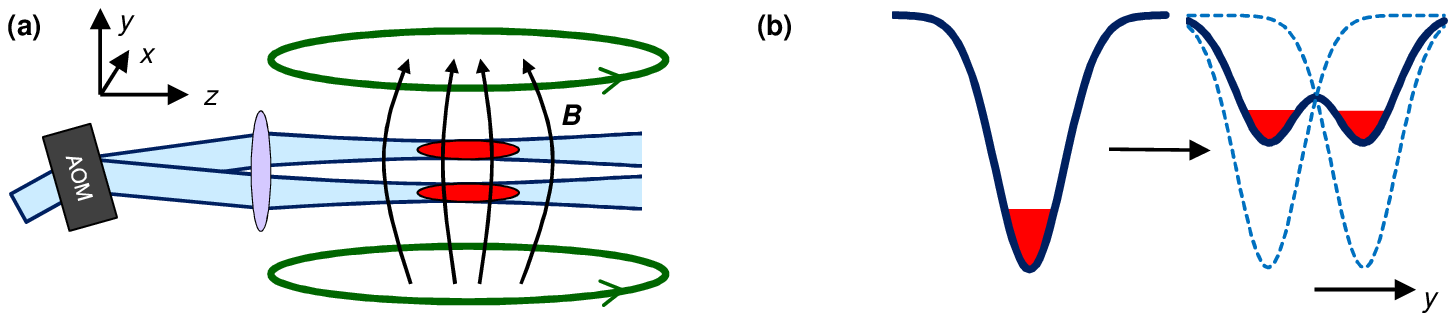}\\
  \caption{
Illustration of the trapping and splitting of the mBEC in the
presence of a magnetic field $B$. An acousto-optical modulator (AOM)
toggles the laser beam between two positions, which creates an
effective double-well potential for trapping two mBECs. (a) Along
the $x$- and $y$-directions, the optical potential is dominant;
along the $z$-axis the magnetic potential is dominant. (b) The
potential shape of the optical dipole trap is Gaussian. The
double-well potential is generated from the superposition of two
Gaussian potentials.
  }\label{figscheme}
  \end{center}
\end{figure}

We create a molecular Bose-Einstein condensate (mBEC), starting from
an atomic Fermi gas consisting of an equal mixture of $^6$Li in the
lowest two spin states. The preparation
follows the procedures described in our previous work
\cite{Jochim2003bec,Bartenstein2004cfa,Altmeyer2007pmo,Riedl2008coo}.

The atoms are trapped in the potential of a focused, far red-detuned
laser beam with a beam waist of $45\,\mu$m, derived from a $25$\,W,
$1030$\,nm single-mode laser source, as illustrated in
Fig.~\ref{figscheme}.
We choose the coordinate system such that the laser beam propagates
along the $z$-axis and gravity acts in $-y$-direction. A magnetic
bias field $B$ can be applied along the $y$-axis. A broad Feshbach
resonance centered at $B = 834$\,G \cite{Bartenstein2005pdo}
facilitates precise tuning of the atomic $s$-wave scattering length
$a$. Below resonance, a weakly bound molecular state exists
\cite{Jochim2003pgo}. Molecules in this state represent halo dimers,
since their wave function extends far into the classically forbidden
range \cite{Ferlaino2008cbt}. Their size is given by $a$ and their
binding energy is $\hbar^2/(ma^2)$, where $m$ denotes the atomic
mass and $\hbar$ is Planck's constant $h$ divided by $2\pi$. The
intermolecular scattering length is $a_M=0.6 a$
\cite{Petrov2005spo}.

To create the mBEC we perform evaporative cooling by reducing the laser beam power at a
constant magnetic field $B=764$\,G. During evaporation, the halo
dimers are created through three-body collisions
\cite{Jochim2003bec} and eventually they form a mBEC
\cite{Inguscio2006ufg}. After evaporation, we increase the trap
depth, thereby compressing the condensate, to avoid spilling
particles in all further steps of the experimental sequence. The
beam power is adiabatically increased by a factor of about $10$ to
$45$\,mW. The trap center can be closely approximated by a harmonic
potential. The oscillation frequencies of the molecules, which are
the same as the ones of free atoms, are $(\omega_x, \omega_y,
\omega_z)=2 \pi \times (250,250,20.6\times \sqrt{B/700\,{\rm
G}})$\,Hz. The axial confinement essentially results from the curvature of the magnetic field.
We obtain a cigar-shaped cloud containing
$N=1.8\times10^5$ molecules. The condensate fraction exceeds
$90\,\%$ \cite{Jochim2003bec}.

Most of our measurements are carried out in the regime of weak
interaction between the molecules. We ramp the magnetic field
adiabatically down to $700$\,G in $200$\,ms, thereby decreasing the
scattering length to about $a_M = 1000\,\rm{a}_0$; at lower fields
the molecules become unstable
\cite{Petrov2005dmi,Cubizolles2003pol,Jochim2003pgo} and limit the
lifetime of the mBEC. At $700$\,G, the chemical potential of the
mBEC is $k_B\times 200$\,nK, with $k_B$ denoting the Boltzmann
constant, and the binding energy of the molecules is $k_B\times
8\,\mu$K.
In view of the crossover from
BEC to a Bardeen-Cooper-Schrieffer (BCS) type regime
\cite{Giorgini2008tou,Inguscio2006ufg}, one can also express the
interaction conditions in terms of the commonly used dimensionless
parameter $1/(k_F a)$, where $k_F$ is the Fermi wave number of a
non-interacting Fermi gas with $(\hbar k_F)^2/(2m) = E_F$, where
$E_F=\hbar (6N \omega_x \omega_y \omega_z)^{1/3}$ is the Fermi
energy. For the condition of our mBEC at $700$\,G we obtain $1/(k_F
a) = 3$. Strongly interacting conditions are realized for $1/(k_F
a)<1$, which can be achieved at fields closer to resonance.

\subsection{Condensate splitting}
\label{sec:splitting}

\begin{figure}
\begin{center}
  \includegraphics[width=6cm]{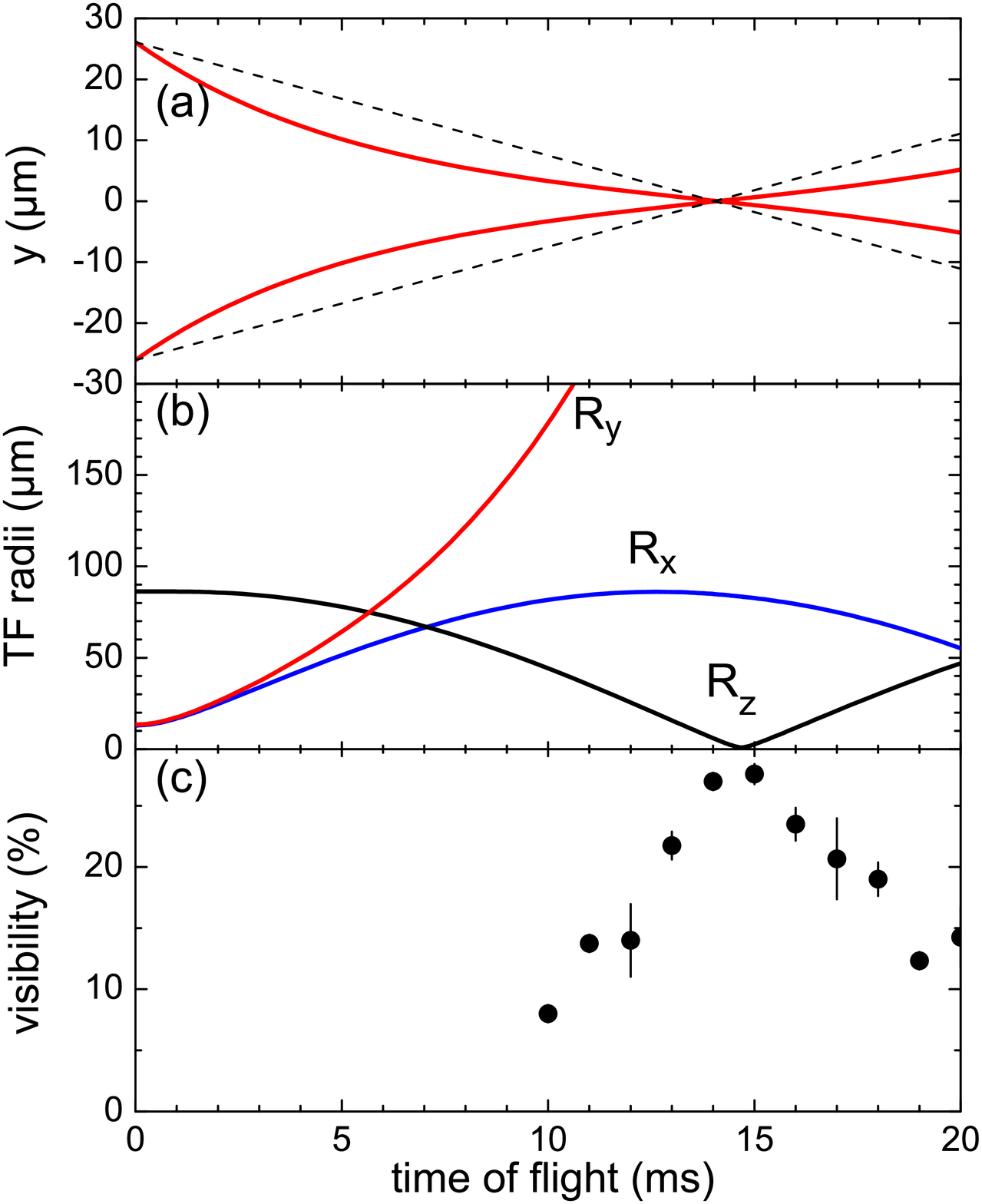}\\
  \caption{
Expansion dynamics of the condensates in the magnetic saddle
potential. (a) The solid lines are the calculated center-of-mass
motion of the condensates, taking into account an initial kick
towards each other, see text. The trajectories intersect after $t_{\rm
TOF}=14$\,ms. For comparison, the dashed lines represent the
trajectories of particles in free expansion intersecting at the same
point. (b) The calculated Thomas-Fermi radii of the
condensates show the expansion along the $x$- and $y$-axis and the
compression along the $z$-axis. The initially cigar-shaped mBEC
evolves into a flat disc. (c) The measured visibility of the fringe
pattern shows a clear peak, which coincides with the minimum in
$R_z$. The bars indicate the statistical uncertainties derived from $10$ individual
measurements.
  }\label{figexpansion}
  \end{center}
\end{figure}

The mBEC is split into two equal parts along the $y$-axis. We
transform the Gaussian shaped optical dipole potential into a
double-well potential, as illustrated in Fig.~\ref{figscheme}(b).
This is accomplished by using time-averaged potentials. An
acousto-optical deflection system modulates the trapping beam
position so fast that the atoms do not follow and feel the
time-averaged beam intensity as their motional potential
\cite{Altmeyer2007doa,Shin2004aiw}. The modulation frequency is
$200$\,kHz and the trapping beam is toggled between two positions,
the distance of which is increased from $0$ to $68\,\mu$m within
$50$\,ms.  The distance between the minima of the resulting double
well is somewhat smaller
because the two Gaussian potentials still overlap. The
measured distance between the centers of the two condensates is
$s=56\,\mu$m and the measured oscillation frequencies in each well
are $(\omega_x,\omega_y, \omega_z)= 2 \pi \times (164,146,20.6\times
\sqrt{B/700\,{\rm G}})$\,Hz. The chemical potential of both
condensates is $k_B\times100$\,nK and the interaction parameter is
$1/(k_F a) = 4$. The barrier height is $k_B\times160$\,nK, which
leads to a fully negligible tunneling rate. The number ratio between
the two condensates after splitting is sensitive to imperfections of
the optical potential. To control equal number splitting, we fine-tune the
magnetic gradient field that is applied to compensate for the
effect of gravity.

\subsection{Expansion in the magnetic field}
\label{sec:expansion}

The specific expansion dynamics of the released mBECs in our setup
is the key to making interference clearly observable, and the
understanding of the expansion is essential for the interpretation
of our results. We identify two effects, which result from the curvature of the magnetic field, that are favorable for the observation of interference.

The coils generating the magnetic offset field in
our set-up are not in Helmholtz configuration, which leads to
second-order terms in $B(x, y, z)$. The resulting magnetic potential
is a saddle potential, where the molecules are trapped along the
$x$- and $z$-directions, but they are anti-trapped along the
$y$-axis, the symmetry axis of the field. The oscillation
frequencies are $(\omega_x, \omega_y,\omega_z)=2 \pi \times
(20.5,i\times 29,20.5)\times \sqrt{B/700\,{\rm G}}$\,Hz, where the
imaginary frequency denotes the anti-trap along the $y$-axis.

We model the expansion by adopting the scaling approach as applied
in Refs. \cite{Menotti2002eoa,Altmeyer2007doa}.
Figure~\ref{figexpansion}(b) shows the predicted evolution of the
Thomas-Fermi (TF) radii $R_x$, $R_y$ and $R_z$, which we also verify experimentally. At the beginning, the
expansion is driven by the pressure gradient in the cloud, which
leads to a fast acceleration in the radial direction. This expansion is
then further accelerated along $y$ and decelerated along $x$ because
of the magnetic saddle potential. Along the $z$-axis, the long axis
of the trapped cloud, the trap remains basically unchanged when the
cloud is released from the optical potential. As the mean field
pressure of the expanding cloud decreases, the magnetic confinement
leads to a spatial compression of the cloud. We find that after
$t_{\rm TOF}\approx 14$\,ms the parameter $R_z$ has a minimum because of this compression effect.

For high interference contrast, large overlap of the two clouds at the time of detection is essential. To achieve this, the condensates are kicked towards each other by
switching on the original single-well trap, typically for $0.1$\,ms
right after release from the double well. The solid lines in
Fig.~\ref{figexpansion}(a) show the calculated center-of-mass motion
of the clouds after the initial kick to assure large overlap at
$t_{\rm TOF}\approx 14$\,ms.

The interference pattern is determined by the relative velocity
between the two condensates. The relative velocity $v_{\rm{rel}}$ at
$y=0$ and $t_{\rm TOF}=14$\,ms can be directly deduced from the
slopes of the solid lines in Fig.~\ref{figexpansion}(a). This
velocity is substantially smaller than it would be in free expansion without
magnetic potential, where particles meeting at $y=0$ and $t_{\rm
TOF}=14$\,ms would follow the dashed trajectories in
Fig.~\ref{figexpansion}(a).
This deceleration of $v_{\rm{rel}}$ can be readily visualized by the
condensates climbing up the potential hill resulting from the
anti-trap in $y$-direction. This anti-trap also accelerates the expansion in $y$-direction, see $R_y$ in Fig.~\ref{figexpansion}(b). Remarkably,  since the velocity field in each of the clouds stays linear, $v_{\rm{rel}}$ is independent of the position. More rigorously, we calculate $v_{\rm{rel}}$ using the scaling approach and taking into account the center-of-mass motion of the clouds.

Thus expansion dynamics brings about two favorable effects: First, the spatial
compression along the $z$-axis facilitates clear detection of
interference fringes by absorption imaging. Second, the decreased
relative velocity leads to an increased fringe period. This means
that the anti-trap acts as a magnifying glass for the interference
fringes.

\subsection{Detection and analysis of interference fringes}
\label{sec:vis}

\begin{figure}
\begin{center}
  \includegraphics[width=10cm]{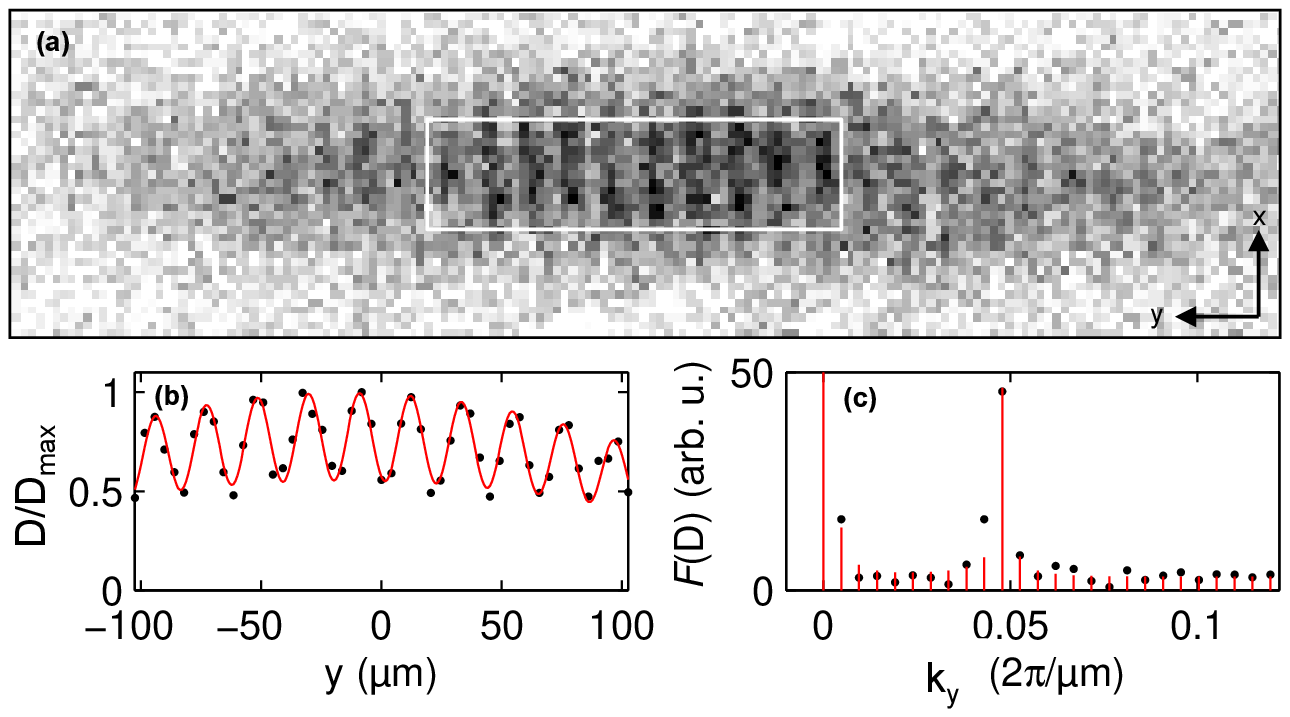}\\
  \caption{
Interference image and analysis. (a) The column density along the
$z$-axis after $t_{\rm TOF}=14$\,ms shows the interference pattern.
The field-of-view is $660\,\mu$m$\times170\,\mu$m. The inner box
indicates the region used for analysis. (b) The column density
integrated along $x$ gives the density distribution $D$ along $y$
(dots). The solid curve is the result of the fit in Fourier space,
see text. (c) The density distribution is Fourier transformed (dots)
and fitted (bars).
  }\label{figinterference}
  \end{center}
\end{figure}

We detect the clouds by absorption imaging.
Figure~\ref{figinterference}(a) shows a typical image of
interference after $14$\,ms time of flight. The imaging beam
propagates along the $z$-axis. It is overlapped with the trapping
beam using dichroic mirrors. The imaging light pulse is on for
$10\,\mu$s and its intensity is about the saturation intensity of $^6$Li
atoms. We state-selectively image the atoms in the second-to-lowest
Zeeman state. Already the first photon scattering event is likely to
dissociate the weakly bound molecule \cite{Bartenstein2004cfa},
followed by about $10$ more photons scattered by the free atom.

From the absorption images, we determine the visibility and fringe
period of the interference pattern. The column density is integrated along
the $x$-direction over the region depicted in
Fig.~\ref{figinterference}(a) \footnote{The size of the region was
chosen to produce the optimal signal to noise.} resulting in a
one-dimensional density distribution $D$, shown in
Fig.~\ref{figinterference}(b). The density distribution contains various kinds of noise (e.g.~photon or atom shot noise, or camera readout noise), which may be misinterpreted as interference signal.
Therefore we analyze the density distribution in Fourier space by considering the Fourier transformed density distribution $ \mathcal{F}(D)$, see
Fig.~\ref{figinterference}(c). Here all those types of noise are approximately
white and show up as a constant offset, whereas, the signal of
interference is monochromatic and shows up as a peak. This gives the
possibility to subtract the average contribution of noise from the
signal. We determine the visibility and fringe period
by the custom fit function in Fourier space
\begin{equation}\label{fitfun}
f=\sqrt{|\mathcal{F}( (a+b~y+c~y^2)\times(1+v~\sin(2
\pi/d~y+\phi)))|^2 +n^2},
\end{equation}
yielding the fringe period $d$, the visibility $v$, and the relative phase
$\phi$. The term $a+b~y+c~y^2$ account for the somewhat non-uniform
density distribution. The white noise $n$ is the offset in Fourier
space. Since the phase between the signal and
the noise is random, the corresponding contributions are added
quadratically. The discrimination of the noise via this fitting routine is crucial when the visibility is low.

The largest observed visibility is about $30\,\%$. We find
that this upper limit can be essentially attributed to the finite resolution of our imaging
system. We determine the modulation transfer function of the imaging system
and it gives about $30\pm10$\,\% visibility for structures with
period $d=20\,\mu$m. Also other sources can contribute to a reduction of visibility, like a blurring because of a limited depth of focus or a tilt of the planes of
constructive and destructive interference. The planes are in general
somewhat tilted with respect to the line of sight, thereby obscuring
the fringe pattern on the image. But these effects are suppressed by the spatial compression
along the imaging axis caused by the magnetic saddle potential. This
can be seen by comparing the compression of $R_z$ in
Fig.~\ref{figexpansion}(b) to the detected visibility in
Fig.~\ref{figexpansion}(c). The minimum of $R_z$ after $t_{{\rm
TOF}}=14$\,ms coincides with the peak in visibility. The peak value
of almost $30$\,\% agrees with the resolution limit of the imaging
system. All following measurements are performed when the clouds are
compressed to about $1\,\mu$m along the imaging axis; in this case, only the
limited resolution is relevant. The spatial compression is an
alternative to the slicing imaging technique used in
Ref.~\cite{Andrews1997ooi} and brings along the advantage that all
particles are imaged.

\section{Experimental results}
\label{sec:res}

The observed interference pattern is the standing wave formed by two
macroscopically occupied matter waves, the two molecular BECs. Here
we present our main experimental results.
In Sec.~\ref{sec:fri}, we investigate the fringe period, which
evidences that the interfering particles are molecules. In
Sec.~\ref{sec:tem}, we study the visibility when heating the cloud
to above the critical temperature for condensation to show that the
interference is established by the condensate fraction. In
Sec.~\ref{sec:ia}, we explore the dependence of the visibility on
the interaction strength and find that non-forward scattering
processes depopulate the momentum component of the matter wave that
is responsible for the interference pattern.

\subsection{Fringe period}
\label{sec:fri}

\begin{figure}
\begin{center}
  \includegraphics[width=8cm]{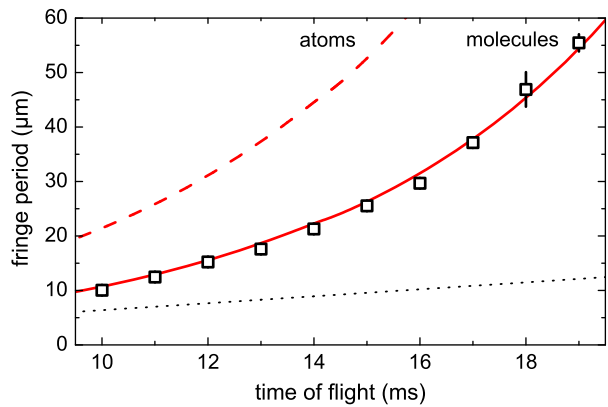}\\
  \caption{
Fringe period as a function of time of flight. The symbols are the
measured periods with bars, mostly smaller than the symbol size,
indicating the statistical uncertainties resulting from $10$
individual measurements at a given time of flight. The solid line is
the calculated period for molecules and the dashed line for atoms.
For free expansion without the magnetic saddle potential, the fringe
period of molecules would be much smaller (dotted line).
  }\label{figmolecules}
  \end{center}
\end{figure}

The fringe period is an central observable in interference
experiments. Figure~\ref{figmolecules} shows the measured fringe
period at $B=700$\,G as a function of time of flight. The de~Broglie
relation yields the fringe period
\begin{equation}
d=\frac{h}{M v_{\rm{rel}}},
\end{equation}
which is determined by the mass $M$ of the interfering particles and
by the relative velocity $v_{\rm{rel}}$ of the two condensates. In
our experiment, we calculate $v_{\rm{rel}}$ from the expansion and
center-of-mass motion of the condensates in the magnetic field
curvature, as discussed in Sec.~\ref{sec:expansion}. The result is
in contrast to the simple relation $v_{\rm{rel}}=s/t_{\rm TOF}$ that
holds for the free expansion usually considered in experiments of
this type.
The solid line in Fig.~\ref{figmolecules} displays the calculated
fringe period $d$ for molecules, where we set $M=2 m$.
All input parameters for this calculation are determined independently. Their combined uncertainties result in typical uncertainty of $3\,\%$ for the fringe period, with the main contribution stemming from the uncertainty in the cloud separation.
The data are
in remarkable agreement with the calculation. For comparison, we also plot the fringe period
for interfering atoms ($M = m$), which is clearly incompatible with
the data.

The dotted line in Fig.~\ref{figmolecules} displays the fringe
period that would result for freely expanding mBECs without the
magnetic saddle potential. Comparing this curve to the much larger
fringe period that we observe, highlights the effect of the magnetic
field curvature to magnify the fringe period, as discussed in
Sec.~\ref{sec:expansion}. The same magnification effect was reported
in Ref.~\cite{Zawadzki2010sif}.

Note that the fringe period can be increased by interaction-induced
slowing down of the two overlapping condensates
\cite{Simsarian2000itp}. The mean-field of one condensate represents
a potential hill for the other condensate, which slows down when
climbing this hill. Under our experimental conditions at $700$\,G,
the effect is found to be negligible. For stronger interaction, we
see indications of this effect in agreement with a corresponding model
calculations.

\subsection{Dependence of interference visibility on condensate fraction}
\label{sec:tem}

\begin{figure}
\begin{center}
  \includegraphics[width=8cm]{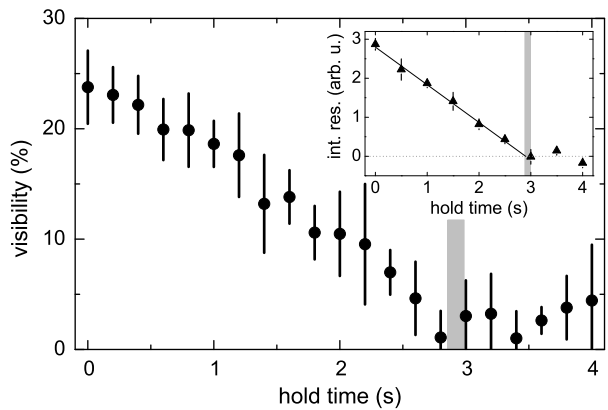}\\
  \caption{
Visibility of interference for increasing temperature. The main
figure shows the measured mean visibility with bars indicating the
standard deviation resulting from $11$ measurements. Here, we plot the standard deviation and not the statistical uncertainty to better illustrate the range of measured values. During the
hold time in the trap, the temperature increases from low
temperature to above $T_c$. The hold time after which $T_c$ is
reached is indicated by the grey bar. The inset shows the integrated
residuals of a Gaussian fit, see text. A linear fit to the first six points facilitates a
simple extrapolation to zero, which marks the vanishing of the
condensate fraction.
  }\label{figT}
  \end{center}
\end{figure}

To demonstrate that the interference results only from the condensed
molecules and not from the thermal fraction, we perform a controlled
heating experiment and show the loss of visibility with vanishing
condensate fraction. Starting from an almost pure condensate
\cite{Jochim2003bec}, we hold the gas in the recompressed optical
dipole trap for a variable hold time before splitting. Intensity
fluctuations and pointing instabilities of the laser beam as well as
inelastic collisions between the molecules \cite{Petrov2005dmi} heat
the gas and lead to a monotonous temperature increase
\cite{Savard1997lni,Wright2007ftc}. To demonstrate that the
interference results from the condensate, it is sufficient to
determine the hold time at which the critical temperature for
condensation $T_c$ is reached. Therefore, we fit a Gaussian profile
to the density distribution of the cloud, which is recorded after
expansion for $t_{\rm TOF}=5$\,ms from the single-well trap. We
find that the integrated residual of the fit gives a good measure whether the
cloud shape deviates from a thermal one.
The inset in Fig.~\ref{figT} shows that the integrated residual goes
to zero after a hold time slightly below $3$\,s, which locates the
phase transition.

The visibility data in Fig.~\ref{figT} are recorded at $B=700$\,G
after $t_{\rm TOF}=14$\,ms \footnote{We verify on images after
$t_{\rm TOF}=0.4$\,ms that the clouds are still separated in the
double-well potential despite the higher thermal energies.}. The
visibility decreases as the temperature increases and vanishes for a
hold time that coincides with the hold time when $T_c$ is reached.
The observed decrease of visibility is continuous because we image
the full column density including the growing thermal fraction, which does not clearly separate from the condensate in expansion at $700$\,G. Above $T_c$, the
density distribution does no more show any fringes. Still, the
fitting routine produces finite mean values because it can output
only positive values. But if the measured visibility is not larger
than the standard deviation, its distinction from zero is not
significant. The vanishing visibility above the critical temperature
confirms that, as expected, the interference is established by the
condensate fraction.

Further intriguing evidence that the interference is caused by the
condensate is the observation of interference between independent
ultracold clouds. An independent production rules out that the
interference can be caused by self interference of particles
\cite{Miller2005hci}. To investigate interference between
independent clouds, we split them already at a temperature far above the
critical temperature to a large distance of $180\,\mu$m and then
create two mBECs independently. Shortly before release, we reduce
the distance to obtain the identical geometry as in all the other
measurements and proceed as usual. We observe the same kind of
interference pattern with a visibility of about $15$\,\%. The lower
visibility can be explained by a less efficient evaporation and less control over the equal number preparation in the double well.

\subsection{Dependence of interference visibility on interaction strength}
\label{sec:ia}

\begin{figure}
\begin{center}
  \includegraphics[width=8cm]{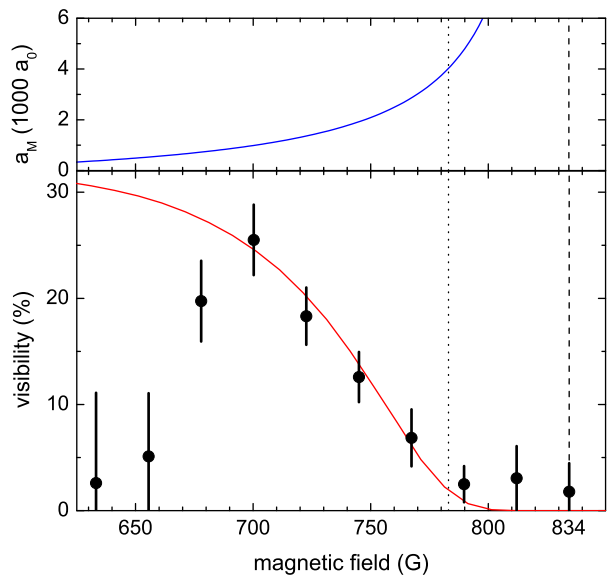}\\
  \caption{
Visibility of interference from weak to strong interaction. The
upper panel shows how the molecular scattering length $a_M$
increases towards the Feshbach resonance at $834$\,G, marked by the
dashed line. The onset of the strongly interacting regime is marked
by the dotted line. In the lower panel, the dots represent the mean
visibility with bars indicating the standard deviation resulting
from $20$ individual measurements.  The solid line is the predicted visibility
from the simple calculation modeling the non-forward scattering
events.
  }\label{figB}
  \end{center}
\end{figure}

In a further set of measurements, we investigate how the fringe
visibility depends on the interaction strength.
Therefore we perform the interference experiment for different
magnetic field values, thereby changing the molecular scattering
length $a_M$ according to the upper panel of Fig.~\ref{figB}
\footnote{We verify on images after $t_{\rm TOF}=0.4$\,ms that the
clouds are still separated in the double-well potential despite the
higher chemical potential at higher interaction strength.}. The
observed visibility as a function of the magnetic field is shown in
the lower panel in Fig.~\ref{figB}. The highest visibility is found
at about $700$\,G. For lower fields, the visibility is decreased,
which we attribute to inelastic decay. The
inelastic collisions of molecules lead to heating of the gas and loss of
particles. The heating reduces the condensate fraction, which
decreases the visibility as observed in the previous section. The
loss also reduces the signal on the images. This leads to a higher
statistical uncertainty in the determination of the visibility,
showing up in the larger standard deviations below $700$\,G.

Towards larger interaction strength, our data show a pronounced
decrease of visibility, and the visibility vanishes at about
$780$\,G. This coincides with the onset of strong interaction
in the trap, where $1/k_Fa\approx 1$. We find that the main effect
causing the decrease is elastic non-forward scattering.
It is known from experimental and theoretical work on colliding
condensates \cite{Chikkatur2000sae,Band2000esl} that elastic
non-forward scattering of particles removes them from the condensate
wave function. In contrast to the forward scattering accounted for
within the usual mean-field approach, this non-forward scattering
transfers particles into momentum states of random direction, which
therefore do no more contribute to the observed interference
pattern. Non-forward scattering is a particle-like excitation, which
requires $v_{\rm{rel}}$ to exceed the speed of sound $v_s$. The
process is suppressed for smaller $v_{\rm{rel}}$
\cite{Chikkatur2000sae,Band2001soe}.  To estimate the decrease of
visibility through this process, we perform a simple model
calculation. The velocity dependence of non-forward scattering is
included by the following approximation: no suppression for
$v_{\rm{rel}}\geq v_s$ and full suppression otherwise. We calculate
the mean number of non-forward scattering events $N_e$ for a
representative molecule with molecules of the other condensate until
the moment of detection. This representative molecule travels along
the center-of-mass path of the condensate; see
Fig.~\ref{figexpansion}(a). We take the bosonically enhanced,
unitarity limited scattering cross section $\sigma=8\pi a_M^2/(1+(k a_M)^2)$,
with $k=m v_{\rm{rel}}/\hbar$. From $N_e$, we derive the
probability for a molecule to still be part of the condensate. This probability is $e^{-N_e}$ and directly corresponds to the
expected visibility, which we fit to the data, excluding the three
data points below $700$\,G. We obtain the solid line in
Fig.~\ref{figB}. The only fit parameter is a normalization factor,
which allows us to account for the reduced detected visibility
because of the limited imaging resolution. The fit yields a factor
of $0.32$, which is consistent with the imaging resolution discussed
in Sec.~\ref{sec:vis}. We find that our simple model for non-forward
scattering can very well explain the decrease of visibility towards
high interaction strength.

There are also other effects that decrease the visibility for
increasing interaction strength, but they turn out to be minor for
our experimental conditions: Strong interaction lead to a depletion
of the condensate \cite{Dalfovo1999tob}. Only the condensate
contributes to the interference pattern and not the depleted
fraction.
The depleted fraction amounts to
about $10\,\%$ at $780$\,G. As we expect the reduction of visibility
to be proportional to the depletion, the reduction is negligible (at
$780$\,G from $2.6$\,\% to $2.3$\,\%). Another effect reducing the
visibility is the collisional dissociation of molecules during
overlap. However, this effect can only occur above $800$\,G, where
the collision energy exceeds the binding energy.

\begin{figure}
\begin{center}
  \includegraphics[width=4cm]{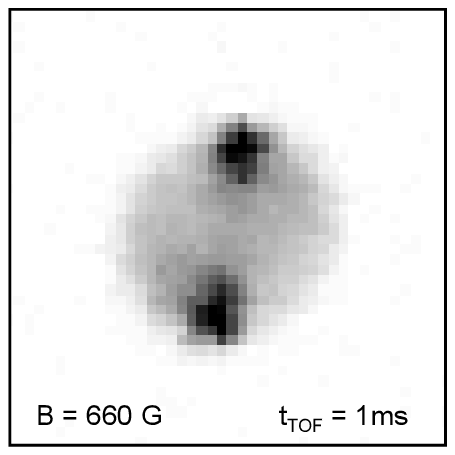}\\
  \caption{
Absorption image $1$\,ms after the collision of two BECs. A
spherical shell of scattered particles clearly separates from the
two BECs. The field of view is $180\times 180\,\mu$m.
  }\label{figcol}
  \end{center}
\end{figure}

To directly demonstrate the effect of non-forward scattering, we study the collision of two
condensates when their relative velocity $v_{\rm{rel}}$ is much
faster than the their expansion velocity. This allows us to observe
the non-forward scattered particles in an $s$-wave shell
\cite{Buggle2004ido}, well separated from the condensates, see
Figure~\ref{figcol}. This separation was not present in the
interference experiments reported before because $v_{\rm{rel}}$ was
similar to the expansion velocity.
We apply our simple model to calculate the fraction of non-forward scattered particles and find good agreement, confirming our model in an independent and direct way.

\begin{figure}
\begin{center}
  \includegraphics[width=12cm]{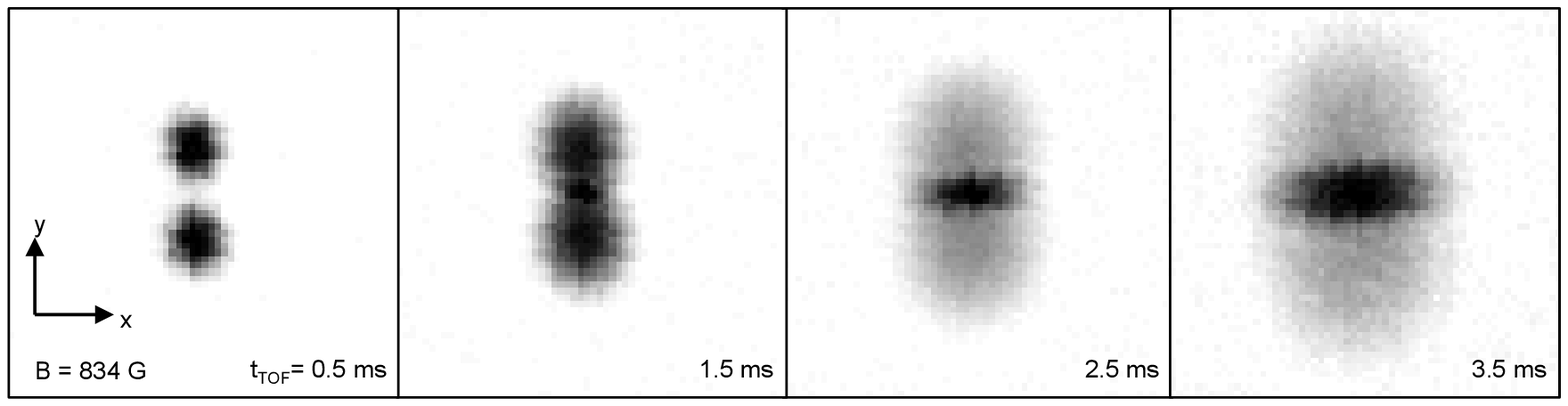}\\
  \caption{
The hindered overlap on resonance. The series shows the first few
milliseconds of expansion. The two clouds do not penetrate each
other, but splash
according to hydrodynamics. The field of view is $180\times
180\,\mu$m.
  }\label{figres}
  \end{center}
\end{figure}

Close to the Feshbach resonance, we enter a regime where the number
of collisions becomes large. This leads to hydrodynamic behavior
also above $T_c$ \cite{Ohara2002ooa,Wright2007ftc}. The time of flight series in
Fig.~\ref{figres}, taken on resonance, shows that the clouds do not
penetrate each other in this regime. Instead, the flow of the particles is
redirected into the the $x$-$z$-plane leading to the observed high
column density in the center. Unlike at low magnetic fields, the
clouds do not superimpose. This directly excludes interference of
two independent condensates in the strongly interacting regime. The
scenario is similar to the one in Ref.~\cite{Joseph2010oos} and
may be described by the analysis therein.

The hindered overlap could be overcome by a magnetic field ramp to
weak interaction after release and before overlapping, as done for
the detection of vortices in Ref.~\cite{Zwierlein2005vas}. Like the
observation of vortices, the observation of interference would
evidence the coherence of the strongly interacting superfluids.

In further measurements, performed above the Feshbach resonance
towards the BCS regime, we did not observe interference. To
discuss possible reasons for the absence of interference fringes, let us first consider the effect of
non-forward scattering on the visibility. As on the BEC side, this effect may hinder overlap and interference for $1/k_Fa<-1$, i.e.~below $910$\,G.
However, we also have to consider that the pairs on the BCS side may
not persist in expansion \cite{Schunck2007seo}, unlike on resonance
or on the BEC side. For the lowest achievable temperature in our
experiment and at $910$\,G, the pairs would be already unstable after a
very short expansion time according to Ref.~\cite{Schunck2007seo}.

\section{Conclusion and outlook}
\label{sec:con}

In conclusion, we have observed the interference between two molecular BECs.
The interference pattern visualizes the standing matter wave of the weakly bound Feshbach molecules and shows coherence over the spatial extension of the cloud. The contrast of interference vanishes above the critical temperature of condensation, demonstrating that the interference is established by the condensed molecules only. We find that non-forward elastic scattering processes can lead to a depletion of the condensate wave function while the clouds overlap. This effect increases towards higher interaction strength and prevents us from observing interference in the strongly interacting regime. On resonance we observe that the two clouds do not overlap but rather collide and deform as a result of deep hydrodynamic behavior.

Interference between condensates of paired fermionic atoms can serve as a powerful tool to investigate many exciting aspects of those systems. A future application will be given, for example, if $p$-wave condensates become available. Here, interference is predicted to reveal the vector nature of the order parameter \cite{Zhang2007mwi}.
A conceptually interesting regime will be entered when the size of
the pairs becomes comparable to the fringe period. Then the detected distribution of atoms may not reveal the interference pattern of the pair distribution. Besides investigating condensates of paired fermions themselves, the system could be used to study the fundamental processes of interference. The wide tunability of the interaction strength could be used to assist self-interference \cite{Cederbaum2007iit} or to investigate to which extent interaction build up the observable relative phase \cite{Xiong2006iii}.

Suppressing the effect of non-forward scattering during overlap could extend the range of applications of condensate interference. Such a suppression may be achieved by reducing the interaction strength before overlap using fast magnetic field ramping techniques \cite{Greiner2003eoa,Zwierlein2005vas}. This technique would allow for investigating the interference in the regime of strong interaction or even on the BCS side of the resonance, where the interference of Cooper-type pairs is an intriguing question in itself.

\section*{Acknowledgments}

We thank Christopher Gaul for stimulating discussions. We acknowledge support by the Austrian Science Fund (FWF) within SFB 15 (project part 21) and SFB 40 (project part 4).

\section*{References}


\providecommand{\newblock}{}

\end{document}